# STORMS PREDICTION: LOGISTIC REGRESSION VS RANDOM FOREST FOR UNBALANCED DATA

**Anne Ruiz-Gazen**

*Institut de Mathématiques de Toulouse and Gremaq, Université Toulouse I, France*

**Nathalie Villa**

*Institut de Mathématiques de Toulouse, Université Toulouse III, France*

*Abstract: The aim of this study is to compare two supervised classification methods on a crucial meteorological problem. The data consist of satellite measurements of cloud systems which are to be classified either in convective or non convective systems. Convective cloud systems correspond to lightning and detecting such systems is of main importance for thunderstorm monitoring and warning. Because the problem is highly unbalanced, we consider specific performance criteria and different strategies. This case study can be used in an advanced course of data mining in order to illustrate the use of logistic regression and random forest on a real data set with unbalanced classes.*

## Introduction

Predicting storms is a high challenge for the French meteorological group Météo France: often, storms are sudden phenomena that can cause high damages and lead to stop some human activities. In particular, airports have to be closed during a storm and planes have to be taken out of their way to land elsewhere: this needs to be planed in order to avoid airports' overbooking or security difficulties. To reach this goal, Météo France uses satellites that keep a watch on the clouds that stand on France; these clouds are named "systems" and several measurements are known about them: some are temperatures, others are morphological measurements (volume, height, ...). Moreover, a "storm network", based on measurements made on earth, can be used to identify, *a posteriori,* systems that have led to storms. Then, a training database has been built: it contains both satellite measurements and information coming from the storm network for each considered system.

The goal of this paper is to present two statistical approaches in order to discriminate the convective systems (clouds that lead to a storm) from the non convective ones: a logistic regression is compared to a random forest algorithm. The main point of this study is that the data set is extremely unbalanced: convective systems represent only 4% of the whole database; therefore, the consequences of such unbalancing and the strategies developed to overcome it are discussed.

The rest of the paper is organized as follows: Section I describes the database. Then, Section II presents the methods and performance criteria that are suitable in this problem. Section III gives some details about the R programs that we used and Section IV presents the results of the study and discusses them.

## I. Description of the data set

The working dataset was divided into two sets in a natural way: the first set corresponds to cloud systems observed between June and August 2004 while the second set corresponds to systems observed during the same months in 2005. So, the 11 803 observations made in 2004 are used to build a training sample and the 20 998 observations made in 2005 are used as a test sample. All the cloud systems are described by 41 numerical variables built from satellite measurements (see the Appendix for details) and by their nature (convective or not). Moreover, the values of the variables have been recorded at 3 distinct moments: all the systems have been observed during at least 30 minutes and a satellite observation is made each 15 minutes. The convective systems variables have been recorded 30 minutes and 15 minutes before the first recorded lightning and at the first lightning moment. A 30 minutes period also leading to three different moments has been randomly sampled by Météo France in the trajectories of the non convective systems. A preprocessing that will not be detailed further has been applied to the data in order to clean them up from highly correlated variables and from outliers.

The reader can find the observations made in 2004 in the file `train.csv` and the observations made in 2005 in the file `test.csv`. The first line of these files contains the names of the variable and the last column ("cv") indicates the nature of the data: 1 is set for "convective" and 0 for "non convective". The same data can also be found in the R file `ruizvilla.Rdata`.

For the rest of the paper, we will denote $X$ the random vector of the $p = 41$ satellite measurements and $Y \in \{0,1\}$ the variable that has to be predicted from *X* (i.e. if the system is convective or not).

As most of the observed systems are non convective, the database is extremely unbalanced: in the training set, 4.06% of the systems are convective and in the test set, only 2.98% of the systems are convective. This very few number of convective systems leads to the problem of training a model from data having a very few number of observations in the class of interest.

## II. Methods

The models that will be developed further lead to estimate the probability $P(Y=1|X)$. The estimate of this conditional probability will be denoted by $\hat{P}(Y=1|X)$. Usually, given these estimates, an observation is classified in class 1 if its estimated probability satisfies

$$\hat{P}(Y=1|X) > 0.5. \quad (1)$$

### A. Unbalanced data set

When the estimation of $P(Y=1|X)$ has been deduced from an unbalanced data set, the methodology described above could lead to poor performance results (see, for instance, Dupret and Koda (2001), Liu et al. (2006)). Several simple strategies are commonly proposed to overcome this difficulty.

The first approach consists in *re-balancing* the dataset. Liu et al. (2006) explore several re-balancing schemes both based on downsampling and on oversampling. Their experiments seem to show that downsampling is a better strategy. Thus, the training set has been re-balanced by keeping all the observations belonging to the minority class (convective systems) and by randomly sampling (without replacement) a given (smaller than the original) number of observations from the majority class (non convective systems). Moreover, as suggested by Dupret and Koda (2001), the optimal re-balancing strategy is not necessarily to downsample the minority class at the same level as the majority class. A previous mining of the dataset (see Section IV, A) leads to choose a downsampling rate equal to:

# Convective / # Non Convective = 0.2.

The sampled database that was used for training the methods can be found in the file `train_sample.csv`

The second approach to treat the problem of unbalanced database is to deduce the final decision from the choice of a *threshold* $\tau$. In equation (1), the threshold is set to 0.5 but, as emphasized for example in Lemmens and Croux (2006), this choice is generally not optimal in the case of unbalanced datasets. Then, the distribution of the estimates $\hat{P}(Y=1|X)$ is studied for all the observations of the training and test samples and the evolution of the performance criterion as a function of the threshold is drawn in order to propose several solutions for the final decision function. This final decision can be adapted to the goal to achieve, making the balance between a good level of recognition for the convective systems and a low level of false alarms.

### B. Description of the methods

We propose to investigate two classification methods for the discrimination of convective systems: the logistic regression and the more recent random forest method. These two approaches have been chosen for their complementary properties: logistic regression is a well-known and simple model based on a generalized linear model. On the contrary, random forest is a non linear and non parametric model combining two recent learning methods: the classification trees and the bagging algorithm. Among all the recent advances in statistical learning, classification trees have the advantage of being very easy to understand, which is an important feature for communication. Combined with a bagging scheme, the generalization ability of such a method has been proven to be very interesting. In the following, we present more deeply the two methods.

*Logistic regression*

"Logistic regression" is a classification parametric model: a linear model is performed to estimate the probability of an observation to belong to a particular class. Using the notations introduced previously, the logistic regression estimates the probability $P(Y=1|X)$ by using the following linear relation and the maximum likelihood estimation method:

$$\ln(P(Y=1|X)/(1-P(Y=1|X))) = \alpha + \sum_{i=1}^{p}\beta_j X_j \quad (2)$$

with $\alpha$ and $\beta_j, j=1,...,p$, the parameters to be estimated. A "step by step" iterative algorithm allows to compare a model based on $p'$ of the $p$ original variables to any of its sub-model (with one less variable) and to any of its top-model (with one more variable): non significant variables are dropped from the following model and relevant ones are added leading to a final model containing a minimal set of relevant variables. This algorithm is performed by the function *stepAIC* from the R library "MASS" and leads to a logistic regression R object denoted by *rl.sbs.opt*. This object can be used for prediction by looking at *rl.sbs.opt$coefficients* which gives the name of the variables used in the final model and their corresponding coefficient $\beta$. The *Intercep* is simply the coefficient $\alpha$ of the model. Using the estimators and computing the model equation for a new system gives a predicted value $S$ that can be transformed into a probability to be convective by:

$$\hat{P} = \exp(S)/(1+\exp(S)).$$

*Random Forest*

"Random Forest" is a classification or regression tool that has been developed recently (Breiman, 2001). It is a combination of tree predictors such that each tree is built independently from the others. The method is easy to understand and has proven its efficiency as a nonlinear tool. Let us describe it more precisely. A classification tree is defined iteratively by a division criterion (node) obtained from one of the variables, $x_j$, which leads to the construction of two subsets in the training sample: one subset contains the observations $i$ that satisfy the condition $x_j^i < T$ while the other subset contains the observations $i$ that satisfy $x_j^i \geq T$ ($T$ is a real number which is defined by the algorithm). The choices of the variable and of the threshold $T$ are automatically built by the algorithm in order to minimize a heterogeneity criterion (the purpose is to obtain two subsets that are the most homogeneous as possible in term of their values of $Y$). The growth of the tree is stopped when all the subsets obtained have homogeneous values of $Y$. But, despite its intuitive form, a single classification tree is often not a very accurate classification function. Thus, Random Forests are an improvement of this classification technique by aggregating several under-efficient classification trees using a "bagging" procedure: at each step of the algorithm, several observations and several variables are randomly chosen and a classification tree is built from this new data set. The final classification decision is obtained by a majority vote law on all the classification trees (and we can also deduce an estimation of the probability of each class by calculating the proportion of each decision on all the classification trees). When the number of trees is large enough, the generalization ability of this algorithm is good. This can be illustrated by the representation of the "out of bag" error which tends to stabilize to a low value: for each observation, the predicted values of the trees that have not been built using this observation, are calculated. These predictions are said "out of bag" because they are based on classifiers (trees) for which the considered observation was not selected by the bagging scheme. Then, by a majority vote law, a classification "out of bag" is obtained and compared to the real class of the observation.

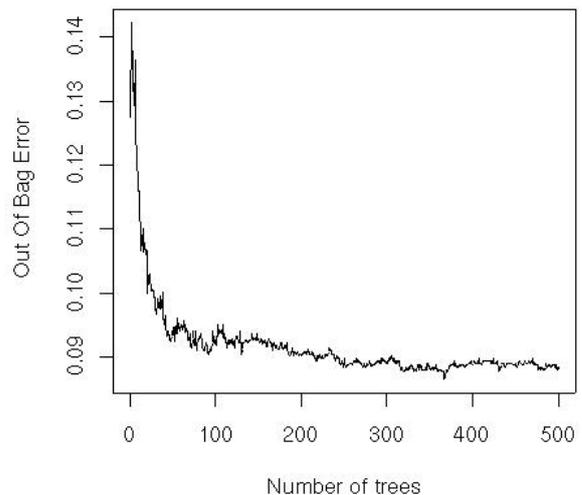

*Figure 1. "Out of bag" error for the training of random forest on the sample* `train.sample`.

Figure 1 illustrates the convergence of the random forest algorithm: it depicts the evolution of the out-of-bag error as a function of the number of trees that was used (for the sample data `train.sample`). We can see that, up to about 300 trees, the error stabilizes which indicates that the obtained random forest has reached its optimal classification error.

C. Performance criteria

For a given threshold $\tau$ (see section II A), the performance of a learning method can be measured by a confusion matrix which gives a cross-classification of the predicted class (Pred) by the true class (True). In this matrix, the number of *hits* is the number of observed convective systems that are predicted as convective, the number of *false alarms* is the number of observed non convective systems that are predicted as convective, the number of *misses* is the number of observed convective systems that are predicted as non convective and the number of *correct rejections* is the number of observed non convective systems that are predicted as non convective.

Usual performance criteria (see for instance, Hand, 1997) are the sensitivity (Se) which is the number of hits over the number of convective systems and the specificity (Sp) which is the number of correct rejections over the number of non convective systems. The well known ROC (receiver operating characteristic) curve plots the sensitivity on the vertical axis against (1 – specificity) on the horizontal axis for a range of threshold values. But in this unbalanced problem, such criteria and curves are not of interest. To illustrate the problem, let us consider the following small example which is not very different from our data set (Table 1).

| True \ Pred | Convective | Non convective | Total |
|---|---|---|---|
| Convective | # hits<br><br>500 | # false alarms<br><br>500 | 1000 |
| Non convective | # misses<br><br>0 | # correct rejections<br><br>9000 | 9000 |
| Total | 500 | 9500 | 10000 |

Table 1

For this example, the sensitivity Se=100% and the specificity Sp≈95% are very high but these good performances are not really satisfactory from the point of view of Météo France. Actually, when looking at the 1000 detected convective systems, only 500 correspond to true convective systems while the other 500 correspond to false alarms; this high rate of false alarms is unacceptable practically as it can lead to close airports for wrong reasons too often.

More generally, because of the unbalanced problem, the number of correct rejections is likely to be very high and so, all the usual performance criteria that take into account the number of correct rejections are likely to give overoptimistic results. So, in this paper, the focus will not be on the ROC curve but rather on the following two performance scores:

- the **false alarm rate (FAR)**:

    FAR= # false alarms /(# hits + # false alarms)

We want a FAR as close to 0 as possible. Note that in the small example, FAR = 50% which is not a good result.

- the **threat score (TS)**:

    TS= # hits /(# hits + # false alarms + # misses)

We want a TS as close to 1 as possible. In the small example, the value of TS=50% which is also not very good.

These performance measures can be calculated for any threshold $\tau$ so that curves can be plotted as will be illustrated in section IV. Note that for a threshold near 0 and because the dataset is unbalanced, the number of hits and the number of misses are expected to be low compared with the number of false alarms and so the FAR is expected to be near 1 and the TS near 0. When the threshold increases, we expect a decrease of the FAR while the TS will increase until a certain threshold value. For high value of threshold, if the number of misses increases, the TS will decrease with the threshold, leading to a value near 0 when the threshold is near 1. These different points will be illustrated in section IV.

## III. Programs

All the programs have been written using R software[1]. Useful information about R programming can be found in R Development Core Team (2005). The programs are divided into several steps:

- a *pre-processing* step where a sampled database is built. This step is performed via the function `rebalance.R` which allows 2 inputs, `train`, which is the data frame of the original training database (file `train.csv`) and `trainr` which is the ratio (# Convective / # Non Convective) to be found after the sampling scheme. The output of this function is a data frame taking into account the ratio `trainr` by keeping all the convective systems found in `train` and by downsampling (without replacement) the non convective systems of `train`. An example of output data is given in file `train_sample.csv` for a sampling ratio of 0.2.

- a *training step* where the training sampled databases are used to train a logistic regression and a random forest.

    The *logistic regression* is performed with a step by step variables selection which leads to find the most relevant variables for this model. The function used is `LogisReg.R` and uses the library "MASS" provided with R programs. It needs 3 inputs: `train.sample` is the sampled data frame, `train` is the original data frame from which `train.sample` has been built and `test` is the test data frame (June to August 2005). It provides the trained model (`rl.sbs.opt`) and also the probability estimates $\hat{P}(Y=1|X)$ for the databases `train.sample`, `train` and `test`, denoted respectively by, `prob.train.sample`, `prob.train` and `prob.test`.

---

[1] Freely available at http://www.r-project.org/

The *random forest* is performed by the function `RF.R` which uses the library "randomForest"[2]. This function has the same inputs and outputs as `RegLogis.R` but the selected model is named `forest.opt`. It also plots the out-of-bag error of the method as a function of the number of trees trained: this plot can help to control if the chosen number of trees is large enough for obtaining a stabilized error rate.

- a *post-processing step* where the performance parameters are calculated. The function `perform.R` has 7 inputs: the probability estimates calculated during the training step (`prob.train.sample`, `prob.train` and `prob.test`), the value of the variable of interest (convective or not) for the three databases and a given value for the threshold, $\tau$, which is named `T.opt`. By the use of the libraries "stats" and "ROCR"[3], the function calculates the confusion matrix for the 3 databases together with the TS and FAR scores associated to the chosen threshold and the Area Under the Curve (AUC) ROC. Moreover, several graphics are built:

  - the estimated densities for the probability estimates, both for the convective systems and the non convective systems;
  - the histograms of the probability estimates both for convective and non convective systems for 3 sets of values ([0,0.2], [0.2,0.5] and [0.5,1]);
  - the graphics of TS and FAR as a function of the threshold together with the graphics of TS versus FAR where the chosen threshold is emphasized as a dot;
  - the ROC curve.

A last function, `TsFarComp.R` creates graphics that compare the respective TS and FAR for two different models. The input parameters of this function are: two series of probability estimates (`prob.test.1` and `prob.test.2`), the respective values of the variable of interest (`conv.test1` and `conv.test2`) and a chosen threshold (`T.opt`). The outputs of this function are the graphics of TS versus FAR for both series of probability estimates where the chosen threshold is emphasized as a dot.

Finally, the programs `Logis.R` and `RandomForest.R` describe a complete procedure where all these functions are used. In `Logis.R` the case where no sampling is made is compared to the case where a sampling with rate 0.2 is pre-processed, for the logistic regression. In `RandomForest.R`, the probability estimates obtained for the re-sampled training set `train.sample` by the use of `forest.opt` are compared to the out-of-bag probability estimates. These programs together with the R functions described above are provided to the reader with the data sets.

## IV. Results and discussion

The previous programs have been used on the training and test databases. In this section, we present some of the results obtained in the study. In particular, we focus on the unbalanced problem and on the comparison between logistic regression and random forest according to different criteria.

### A. Balancing the database

In this section, we consider the logistic regression method and propose to compare the results for different sampling rates in the non convective training database. First, let us consider the original training database containing all the recorded non convective systems from June to August 2004. A stepwise logistic regression is carried out on this training database as detailed in the previous section (program RegLogis) and predicted probabilities are calculated for the training and for the test databases. As already mentioned, once the predicted probabilities were calculated, the classification decision depends on the choice of a threshold $\tau$ which is crucial.

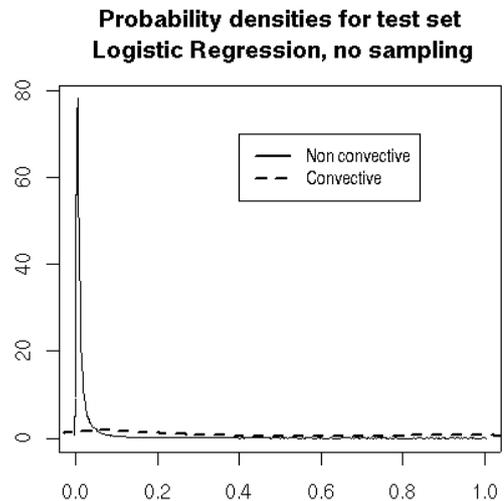

*Figure 2. Densities of the predicted probabilities of the test set for logistic regression trained on* `train`.

Figure 2 plots kernel density estimates of the predicted probabilities for the non convective (plain line) and the convective (dotted line) cloud systems for the test database using the `train` set for training. The two densities look very different with a clear mode for the non convective systems near 0. This could appear as an encouraging result. However, results are not very good when looking at the FAR and TS.

---

[2] L. Breiman and A. Cutler, available at http://stat-www.berkeley.edu/users/breiman/RandomForests

[3] T. Sing, O. Sander, N. Beerenwinkel and T. Lengauer, available at http://rocr.bioinf.mpi-sb.mpg.de

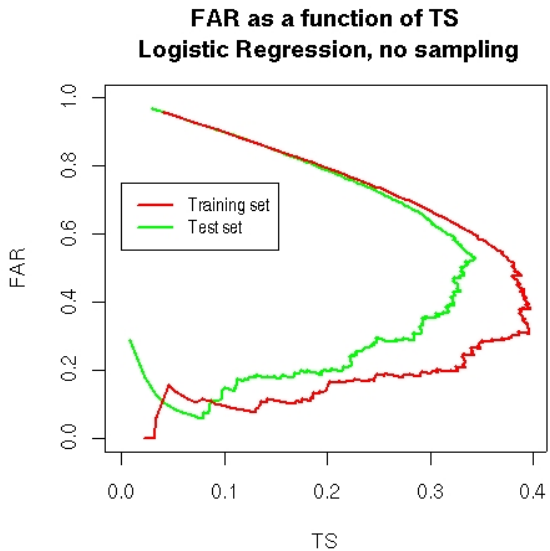

*Figure 3. FAR as a function of TS for logistic regression for the training set (green) and the test set (red) using the training set* `train.`

Figure 3 plots the FAR as a function of TS for different threshold values from 0 to 1 for the training database (in green) and for the test database (in red). As expected (see section II), for low values of the threshold which correspond to high values of the FAR, the FAR decreases when the TS increases. On the contrary, for low values of the FAR (which are the interesting ones), the FAR increases when TS increases. As a consequence, the threshold associated with the extreme point on the right of the figure is of interest only if it corresponds to an admissible FAR. On Figure 3, such a particular threshold is associated with a FAR ≈ 20% and a TS ≈ 35% for the training database (which is not too bad). But it leads to a FAR ≈ 40% and a TS ≈ 30% for the test database. A FAR of 40% is clearly too high according to Météo France objective but, unfortunately, in order to obtain a FAR of 20%, we should accept a lower TS (≈ 20%).

Let us now compare the previous results with the ones obtained when considering a sample of the non convective systems. Focus is on the test database only. The sampling rate is taken such that the number of non convective systems is equal to five times the number of convective systems (sampling ratio = 0.2).

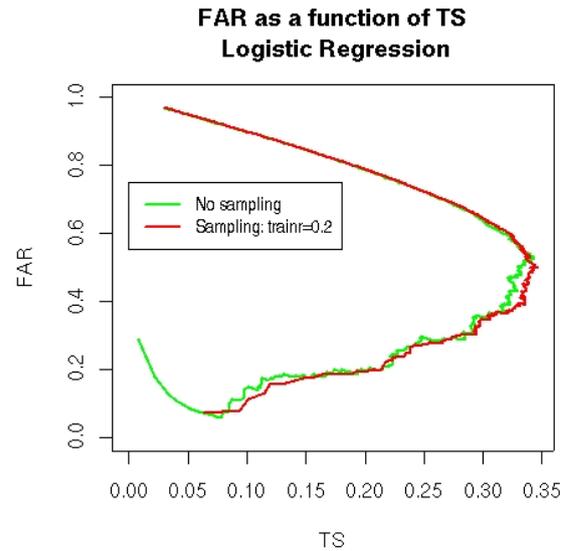

*Figure 4. FAR as a function of TS for the test set for logistic regression trained with* `train` *(green) or* `train.sample` *(red) with 500 discretization points.*

When looking at Figure 4, the TS/FAR curves look very similar. When sampling, there is no loss in terms of FAR and TS but there is no benefit either. In other words, if a threshold $\tau_1 \in [0,1]$ is fixed when taking into account all the non convective systems in the training database, $\tau_1$ corresponds to some FAR and TS; it is always possible to find a threshold $\tau_2$ leading to the same FAR and TS for the test database when training on the sampled database. So, the main advantage of sampling is that because the training data base is much smaller, the classification procedure is much faster (especially the stepwise logistic regression) for similar performance.

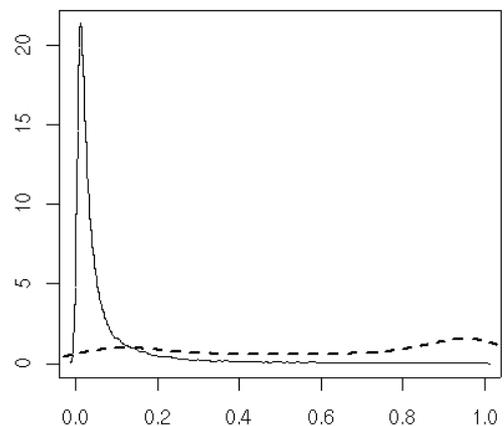

*Figure 5. Densities of the probability estimates of the test set for logistic regression trained on* `train.sample`.

Figure 5 gives the density of the predicted probabilities for the non convective (plain line) and the convective (dotted line) cloud systems for the test database using the `train.sample` set for training. Comparing Figures 2 and 5 is not informative because although they are quite different, they correspond to very similar performances in terms of the TS and the FAR.

Note that further analysis shows that the sampling ratio 0.2 seems a good compromise when considering computational speed and performance. Actually, the performance in terms of the FAR and the TS is almost identical when considering larger sampling rates while the performance decreases a bit for more balanced training databases. As illustrated by Figure 6 which gives a comparison of FAR/TS curves for a sampling ratio of 0.2 and 1, the performance criteria are a little bit worse when the training database is completely rebalanced. Also note that, for the random forest algorithm, very similar results were obtained and thus will not be detailed further.

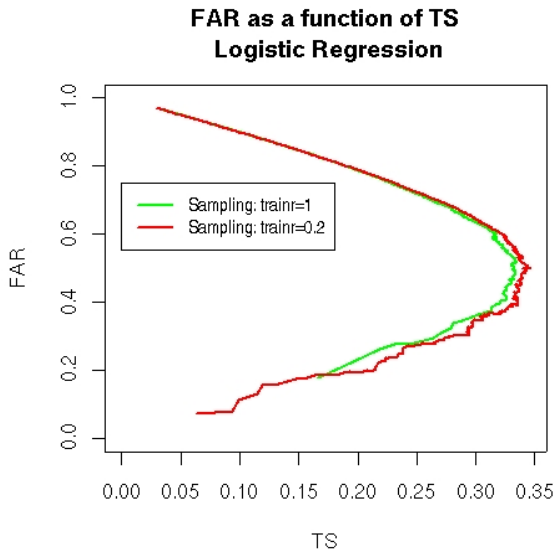

*Figure 6. FAR as a function of TS for logistic regression on `train.sample` (red) and on a training sample with a `trainr = 1` (re-balanced) with 500 discretization points.*

B. Random forest versus logistic regression

Here, as explained in the previous section, we focus on the sampled database `train.sample` for which the sampling ratio (0.2) appears to be a good choice for both logistic regression and random forest. We now intend to compare the two methodologies and explain their advantages and drawbacks.

Let us first focus on the comparison of the densities of the probability estimates obtained by both models. Figures 7 and 9 illustrate this comparison for the training (`train.sample`) and the test sets. The first salient thing is the difference between direct calculation of the probability estimates for the random forest model and the out-of-bag calculation for the training set: Figure 7 shows distinct densities, both for convective and non convective systems (black and green curves).

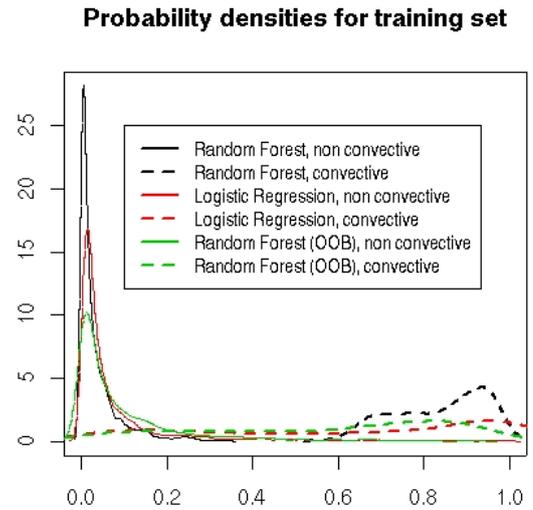

*Figure 7. Densities of the probability estimates for both random forest and logistic regression on the training set.*

The probability estimates for the random forest method are given for "out-of-bag" values (green) and for direct estimation from the final model (black).

This phenomenon is better explained when looking at the respective ROC curves obtained for random forest from the probability estimates calculated from the whole model and those calculated by the out-of-bag scheme (Figure 8).

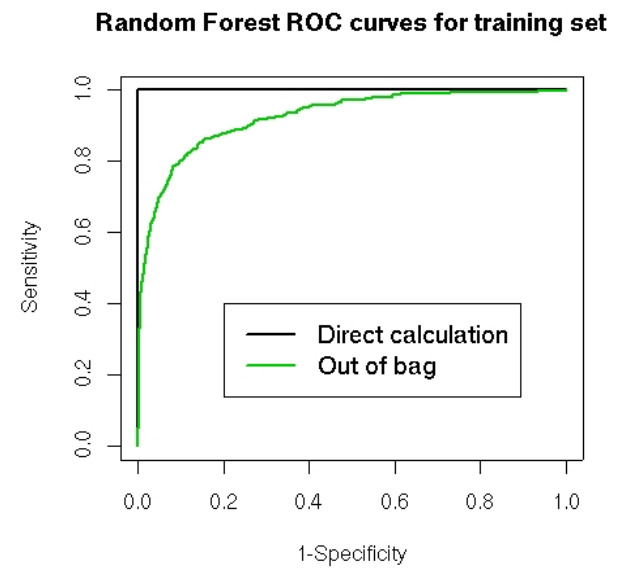

*Figure 8. ROC curves for Random Forest for the training set built from the probability estimates directly calculated by the use of the chosen model (black) and from the out-of-bag probability estimates (green).*

This figure shows that the ROC curve built from the direct calculation of the probability estimates (black) is almost perfect whereas the one obtained from the "out-of-bag" probability estimates is much worse. This illustrates well that random forests overfit the training set, giving an almost perfect prediction for its observations. As a consequence, the probability densities built directly from the training set (black in Figure 7) have to be interpreted with care: although they seem to look better than the logistic regression ones, they could be misleading about the respective performances of the two models.

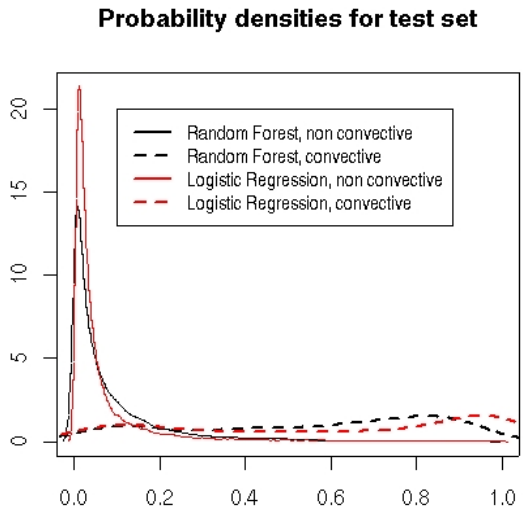

*Figure 9. Densities of the probability estimates on the test set for random forest and logistic regression.*

Moreover, looking at Figure 9, logistic regression seems to better discriminate the two families of probability estimates (for convective systems and non convective systems) than the random forest method for the test set. But, this conclusion should also be taken with care, due to possible scaling differences.

To provide a fairer comparison, the graph of FAR versus TS is given in Figure 10 for both random forest (black) and logistic regression (red). The first remark from this graphic is that the performances of both models are rather poor: for a FAR of 30%, which seems to be a maximum for Météo France, the optimal TS is about 25-30% only.

Focusing on the comparison between random forest and logistic regression, things are not as easy as the comparison of probability densities (Figure 9) could leave to believe. Actually, the optimal model strongly depends on the objectives: for a low FAR (between 15-20%), the optimal model (that has the highest TS) is logistic regression. But if higher FAR is allowed (20-30%), then random forest should be used for prediction.

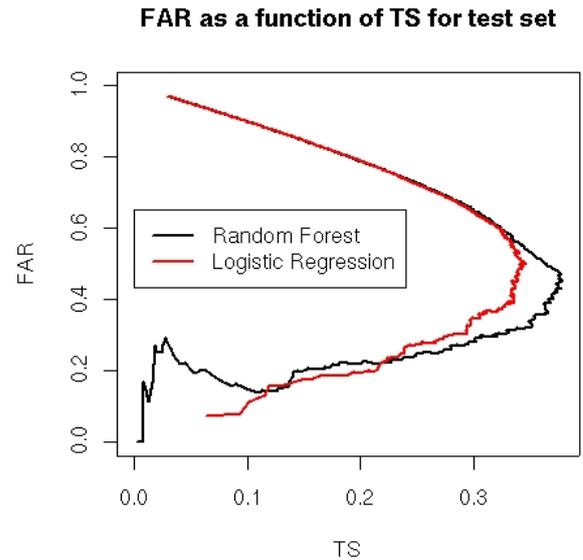

*Figure 10. FAR as a function of TS for the test set (500 discretization points).*

These conclusions give the idea that the two sets of densities can be simply re-scaled one from the other (approximately). Figure 11 explains in details what happens: in this figure, the density of the probability estimates for the logistic regression is estimated, as in Figure 7, but the densities for the random forest method have been re-scaled.

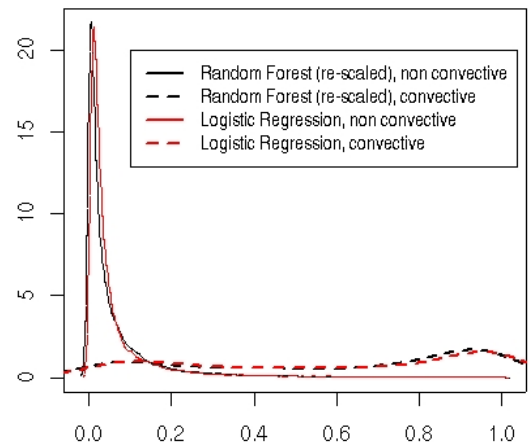

*Figure 11. Rescaled probability densities for the test set (details about its definition are given in the text).*

More precisely, these probability estimates have been transformed by the following function:

$$\varphi(x) = \begin{cases} \alpha_x x & \text{if } x \leq 0.5 \\ 1 - \beta_x(1-x) & \text{if } x > 0.5 \end{cases}$$

where $\alpha_x = 2(1-0.6)x + 0.6$ and

$$\beta_x = 2(1-10^{-3})(1-x) + 10^{-3}.$$

Clearly, this simple transformation does not affect the classification ability of the method (the ranking of the probabilities is not modified) and shows that random forest and logistic regression have very close ability to discriminate the two populations of clouds. Moreover, a single rank test (paired Kendall test) proves that the ranking of the two methods are strongly similar, with a p-value of this test is less than 1E-15.

## V. Conclusion

Several conclusions can be driven from this study: the first one is that unbalanced data lead to specific problems. As was proven in this paper, results obtained with this kind of data should be taken with care. Usual misclassification errors may be irrelevant in this situation: ROC curves can lead to overoptimistic interpretations while comparison of predicted probability densities can lead to misunderstanding. In this study, where a low false alarm rate is one of the main objectives, curves of FAR versus TS are particularly interesting for interpretation and comparisons.

Another interesting aspect of this study is the surprisingly comparable results of the simple logistic regression compared to the popular and more recently developed random forest method. Hand (2006) already argued that the most recent classification methods may not improve the results over standard methods. He points out the fact that, especially in the case of mislabellings in the database, the gain made by recent and sophisticated statistical methods can be marginal. Both, the way the database has been built (by merging satellites sources and storm network records) and the poor performance results obtained with several classification methods, tend to suspect mislabellings. As the logistic regression is a faster method than the random forest and because it leads to simple interpretation of the explanatory variables, it can be of interest for meteorologists.

Finally, to better analyse the performances of the proposed methods, more information could be added to the database. For instance, a storm proximity index could indicate if false positive systems are closer to a storm than true positive ones.

## Acknowledgments

We would like to thank Thibault Laurent and Houcine Zaghdoudi for cleaning the original database and helping us in the practical aspects of this study. We are also grateful to Yann Guillou and Frédéric Autones, engineers at Météo France, Toulouse, for giving us the opportunity to work on this project and for providing the databases.

Correspondence: ruiz@cict.fr (Anne Ruiz-Gazen) and nathalie.villa@math.ups-tlse.fr (Nathalie Villa)

## Appendix: variables definition

In this section, we describe the explanatory variables that were collected by satellite in order to explain storms. In the following, *T* will design the instant of the storm (for convective systems) or the last sampled instant (for non convective systems). $\Delta t$ will be the difference of time between two consecutive records (15 minutes).

### Temperature variables

- `TsBT.0.0` is the threshold temperature at *T* recorded at the basis of the cloud tower.
- `toTmoyBT.0.0` is the mean rate of variation of the mean temperature between *T* and *T*-$\Delta t$ at the basis of the cloud tower.
- `Tmin.0.30` is the minimal temperature at *T*-2$\Delta t$ recorded at the top of the cloud tower.
- `toTmin.0.0`, `toTmin.0.15` and `toTmin.0.30` are the minimal temperatures at *T* recorded at the top of the cloud tower (`toTmin.0.0`) or the mean rate of variation of the minimal temperature between *T* and *T*-$\Delta t$ (`toTmin.0.15`) or between *T*-$\Delta t$ and *T*-2$\Delta t$ (`toTmin.0.30`) recorded at the top of the tower of the cloud.
- `stTmoyTminBT.0.0` and `stTmoyTminBT.0.15` are the differences between the mean temperature at the basis of the cloud tower and the minimal temperature at the top of the cloud tower at *T* (`stTmoyTminBT.0.0`) or at *T*-$\Delta t$ (`stTmoyTminBT.0.15`).
- `stTmoyTminST.0.0`, `stTmoyTminST.0.15` and `stTmoyTminST.0.30` are the differences between mean temperature and the minimal temperature at the top of the cloud tower at *T* (`stTmoyTminST.0.0`), at *T*-$\Delta t$ (`stTmoyTminST.0.15`) or at *T*-2$\Delta t$ (`stTmoyTminST.0.30`).
- `stTsTmoyBT.0.0`, `stTsTmoyBT.0.15` and `stTsTmoyBT.0.30` are the differences between the mean temperature and the threshold temperature at the basis of the cloud tower at *T* (`stTsTmoyBT.0.0`), at *T*-$\Delta t$ (`stTsTmoyBT.0.15`) or at *T*-2$\Delta t$ (`stTsTmoyBT.0.30`).
- `stTsTmoyST.0.0`, `stTsTmoyST.0.15` and `stTsTmoyST.0.30` are the differences between the mean temperature at the top of the cloud tower and the threshold temperature at the basis of the cloud tower at *T* (`stTsTmoyST.0.0`), at *T*-$\Delta t$ (`stTsTmoyST.0.15`) or at *T*-2$\Delta t$ (`stTsTmoyST.0.30`).

### Morphological variables

- `Qgp95BT.0.0`, `Qgp95BT.0.15` and `Qgp95BT.0.30` are 95% quantiles of the gradients (degrees per kilometer) at the basis of the cloud tower at *T*, *T*-$\Delta t$ and *T*-2$\Delta t$.
- `Qgp95BT.0.0.15` and `Qgp95BT.0.15.30` are the change rates between `Qgp95BT.0.0` and `Qgp95BT.0.15` and between `Qgp95BT.0.15` and `Qgp95BT.0.30` (calculated at the preprocessing stage, from the original variable).
- `Gsp95ST.0.0.15` and `Gsp95ST.0.15.30` are the change rates between the 95% quantile of the gradients (degrees per kilometer) at the top of the cloud tower between *T* and *T*-$\Delta t$ and between *T*-$\Delta t$ and *T*-2$\Delta t$ (calculated, as a pre-processing, from the original variables).
- `VtproT.0.0` is the volume of the cloud at *T*.
- `VtproT.0.0.15` and `VtproT.0.15.30` are the change rates of the volumes of the cloud between *T* and *T*-$\Delta t$ and between *T*-$\Delta t$ and *T*-2$\Delta t$ (calculated, as a pre-processing, from the original variables).
- `RdaBT.0.0`, `RdaBT.0.15` and `RdaBT.0.30` are the "aspect ratios" (ratio between the largest axis and the smallest axis of the ellipse that models the cloud) at *T*, *T*-$\Delta t$ and *T*-2$\Delta t$.
- `RdaBT.0.0.15` and `RdaBT.0.15.30` are the change rates of the aspect ratio between *T* and *T*-$\Delta t$ and between *T*-$\Delta t$ and *T*-2$\Delta t$ (calculated, at the preprocessing stage, from the original variables).
- `SBT.0.0` and `SBT.0.30` are the surfaces of the basis of the cloud tower at *T* and *T*-2$\Delta t$.
- `SBT.0.0.15` and `SBT.0.15.30` are the change rates of the surfaces of the basis of the cloud tower between *T* and *T*-$\Delta t$ and between *T*-$\Delta t$ and *T*-2$\Delta t$ (calculated, at the preprocessing stage, from the original variables).
- `SST.0.0`, `SST.0.15` and `SST.0.30` are the surfaces of the top of the cloud tower at *T*, *T*-$\Delta t$ and *T*-2$\Delta t$.
- `SST.0.0.15` and `SST.0.15.30` are the change rates of the surfaces of the top of the cloud tower between *T* and *T*-$\Delta t$ and between *T*-$\Delta t$ and *T*-2$\Delta t$ (calculated, at the preprocessing stage, from the original variables).